\documentclass[twocolumn,secnumarabic,amssymb, nobibnotes, aps, prd]{revtex4-1}

\usepackage{graphicx}

\newcommand{\be}{\begin{equation}}
\newcommand{\ee}{\end{equation}}
\newcommand{\bea}{\begin{eqnarray}}
\newcommand{\eea}{\end{eqnarray}}

\begin{document}

\preprint{AIP/123-QED}

\title{Evidence for A Two-dimensional Quantum Wigner Solid in Zero Magnetic Field}
\thanks{email:jianhuang@wayne.edu}

\author{Jian Huang}
\affiliation{%
Department of Physics and Astronomy, Wayne State University, Detroit, MI 48201, USA\\}%
\author{L. N. Pfeiffer}%
\author{K. W. West}%
\affiliation{%
Department of Electrical Engineering, Princeton University, Princeton, NJ 08544}%

\date{\today}

\begin{abstract}
{We report the first experimental observation of a characteristic nonlinear threshold behavior from dc dynamical response as an evidence for a Wigner crystallization in high-purity GaAs 2D hole systems in zero magnetic field. The system under increasing current drive exhibits voltage oscillations with negative differential resistance. They confirm the coexistence of a moving crystal along with striped edge states as observed for electrons on helium surfaces. However, the threshold is well below the typical classical levels due to a different pinning and depinning mechanism that is possibly related to a quantum process.}
\end{abstract}

\pacs{Valid PACS appear here}
\maketitle

In strongly correlated many electron systems, remarkable manifestations of quantum physics emerge in response to strong inter-particle Coulomb energy ($E_C$). The Fractional Quantum Hall (FQH)~\cite{Tsui} state (with filling factors of odd and even denominators) and Mott insulators~\cite{Mott-VRH} are good examples. The most prominent interaction-driven effect is the Wigner crystallization (WC)~\cite{wc} of electrons which is a solid phase of spatially separated charges settling in a form of a lattice. Such a fascinating quantum state of matter (with spin ordering) can be utilized for futuristic applications such as quantum electronics and spintronics. The classical version of the crystallization, with the Debye temperature $\Theta_D < E_C$, has been demonstrated with 2D electrons on helium surfaces (EHS)~\cite{GrimesAdams,WC_helium}. On the other hand, the more desired quantum version with the Fermi energy $E_F\ll E_C\ll \Theta_D$ has not been previously observed in 2D systems in zero magnetic field. This letter includes experimental evidence of a quantum WC obtained via an ultra-sensitive dc transport study.

The nonlinear transport behaviors~\cite{Kono} in the absence of the significant effect of heating, as demonstrated in both  WC in EHS and the charge density waves (CDW), are evidence for the sliding of spatially ordered charges under bias. Specifically, the threshold/switching behaviors accompanied by resistivity oscillations highlights a characteristic signature for the pinning and depinning of a classical WC~\cite{WC_helium}. However, in order to capture a very delicate quantum WC, much more stringent requirements have to be met with respect to both the high-resolution of dc-VI transport measurement and, especially, the high-purity of the systems. First, interaction effect must be dominant which means the interaction parameter $r_s=a/a_B$, a ratio of the Coulomb energy ($E_C=e^2/(\epsilon\cdot2a$) and the Fermi energy ($E_F=(\pi\hbar^2/m^*)p$), has to be at least 37~\cite{wc1}. In a zero magnetic field, it can only be realized with very dilute charge concentrations which is difficult to achieve in bulk materials. $a=1/\sqrt{\pi p}$ is half of the average charge spacing, $a_B=\hbar^2\epsilon/m^* e^2$-Bohr radius, and $p$-charge density.

Second, the dilute charge density means an enormous average carrier separation, $\propto n^{-1/2}$, which, in the presence of a usual level of disorder, exceeds the single-particle localization length $\xi$. Consequently, interaction effects are overshadowed by the single-particle localization~\cite{Anderson'58} and the system becomes a Wigner glass (distorted/defected WC) or Anderson Insulator. The activated hopping conductance $\sigma\sim e^{-(T^*/T)^\nu}$, as also demonstrated in the insulating side of the metal-to-insulator transition(MIT)~\cite{mit}, is proof of this disorder-driven effect and understood under the framework of Anderson localization. [$T^*$-activation energy; $\nu=1$ is for Arrhenius; $\nu=1/3$ and 1/2 for variable range hoppings (VRH)\cite{Mott-VRH,ES}]. Therefore, a systems with minimal disorder is required so that interaction remains dominant even for $r_s>37$.

We note that the rigorous requirement for the low disorder can be alleviated in the fractional quantum Hall regime in a strong magnetic field where peculiar insulating phases characterized also by activated conductance, near filling factors $1/5$, $1/3$, and 1, have been considered as candidates for a WC~\cite{FWC,reentrant-hi,zhu,xuan}. However, the electron wavefunctions and interaction potentials are radically modified by the large field and the many-electron states are drastically altered. The definitive correlation between the reentrant insulating states and WC is not entirely clear since other candidates are present. The wavefunctions in the zero field, on the other hand, are fully preserved.

Third, the small magnitude of $E_C$ as a consequence of the low carrier density results in a low melting point ($T_m$) of a WC close to experimental temperatures. Moreover, quantum fluctuations as well as residual disorder effect further lower $T_m$. This requires an experiment with effective low temperature cooling. Fourth, the pinning threshold due to weak disorder could be much smaller than expected even though the resistivity ($\rho$) could be enormously large. An electrometer level measurement setup with ultrahigh resolution is crucial.
%

We present a study that meets all four of the above criteria. We adopt 2D hole systems in undoped GaAs/AlGaAs field-effect-transistors called HIGFETs~\cite{lilly,noh,jian-1} that have significantly suppressed disorder~\cite{jian-image}. The larger hole effective mass, $m^*\sim0.3-0.45m_0$~\cite{mass}, means an enhanced interaction effect by several folds compared to electrons ($m^*=0.067m_0$). It has been demonstrated that the 2D concentration can be tuned to a record low value of $6\times10^{8}$ cm$^{-2}$~\cite{jian-1}, with an $r_s>40$. Remarkably, only nonactivated power-law conductance~\cite{jian-1} has been observed even for largest $r_s$ values, indicating an interaction-driven state~\cite{jian-image} instead of an Anderson Insulator.

We report dc measurement results for a dilute 2D holes system, with $p=2.8\times10^{9}$ cm$^{-2}$ and $r_s\sim45$ by assuming $m^*=0.3m_0$~\cite{mass}, in a {\it p}-channel GaAs HIGFET. Contrast to the EHS case, $\Theta_D$ is $\sim 300$ K which greatly exceeds $E_C\sim15$ K, making it a true quantum scenario. The dc measurement is made with a current excitation within $I_{drive}\leq 40 p$A with a remarkable signal resolution of 0.1 $p$A and 0.1 $\mu$V, and performed at various fixed temperatures between 29 mK and 43 mK, a range that is less than the classical $T_m\sim0.225\times10^{-6}p^{1/2}\sim120$ mK. For $T<40m$K, we have observed a threshold behavior closely resembling those found for the EHS~\cite{GrimesAdams,WC_helium} and CDWs~\cite{CDWreview}, even though the threshold is only one thousandth to one millionth of those for EHS and CDWs respectively. Moreover, voltage oscillations with negative differential resistance (NDR), a feature that is almost identical to the EHS~\cite{WC_helium}, is also observed as a function of $I_{drive}$ just above the threshold. The T-dependence of the amplitudes of the oscillations is consistent with a WC melting well below the classical $T_m$. Moreover, comparison drawn with the EHS, CDWs, Si-MOSFETs~\cite{Kravchenko}, and GaAs 2D holes~\cite{Yoon} highlights a strikingly small depinning threshold $I_C$ that could be an indication of certain quantum processes in the pinning/depinning mechanism inspired by an early theory for the CDW studies~\cite{Bardeen,Maki,Rozhavsky,Miller}.\\

\begin{figure}
\includegraphics[totalheight=0.85in,trim=0.2in 0.10in 0.20in 0in]{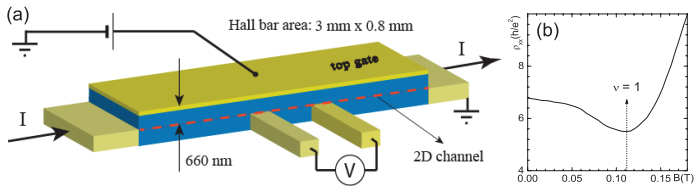}
\caption{\label{fig:hall} (a) schematics for the Hall bar sample. (b) $\rho_{xx}(B)$ showing filling factor 1 at 0.112 Tesla.}
\end{figure}

The device geometry is a standard 3\,mm$\times 0.8$\,mm Hall bar [Fig.~\ref{fig:hall}(a)]. The sample preparation details are provided in Ref.~\cite{jian-fab}. The 2D holes are capacitively induced via a metallic gate that is 660 nm away and the charge density of $2.8\times10^{9}$ cm$^{-2}$ is determined through measuring the magnetoresistance $\rho_{xx}(B)$ [Fig.~\ref{fig:hall}(b)]. Since $a=1/\sqrt{\pi p}\sim$110 nm which is 1/6 of the distance to the top gate, the screening effect from the gate is insignificant. The T-dependence of $\rho(T)$, not shown here, from the ac lock-in measurement in zero magnetic field shows a critical density of $p_c\sim5.5\times10^{9}$ cm$^{-2}$ for the apparent MIT, and is qualitatively nonactivated in the insulating side~\cite{jian-image,jian-1}.

\begin{figure}
\vspace{-0pt}
\includegraphics[totalheight=4.1in,trim=0.37in 0.10in 0.20in 0in]{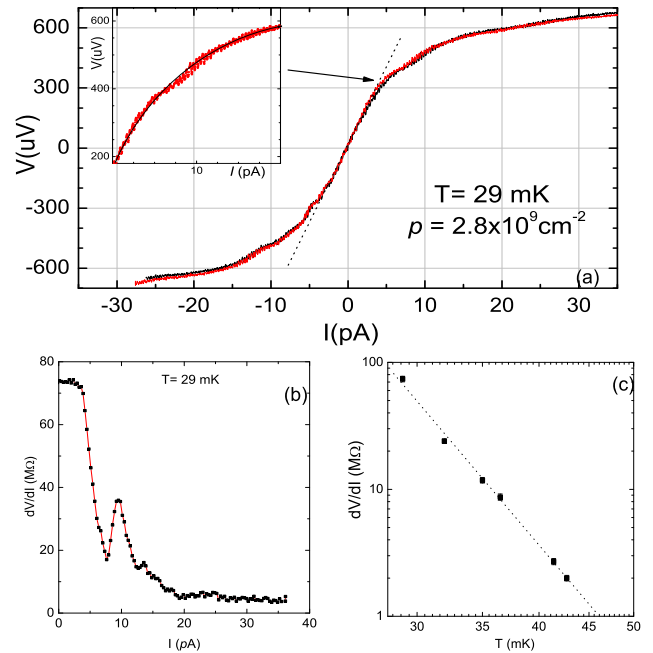}
\vspace{-0pt}
\caption{\label{fig:29mK} (a) Nonlinear dc $V-I$ characteristics for $p=2.8\times10^{9}$ cm$^{-2}$ measured at 29.3 mK. inset: A zoom-in view of the marked area; the solid line is a fit to $I-I_c=\alpha(V-V_c)^{\xi}$. (b) Differential resistance $dV/dI$ vs. $I_{drive}$. (c) log-log scaled T-dependence of the differential resistance $dV/dI$ corresponding to the linear portion below the threshold for various T values.}
\vspace{-0pt}
\end{figure}

Fig.\ref{fig:29mK}(a) shows two dc $V-I$ curves, almost perfectly overlapped, measured at $29.3$ mK for up and down current sweeps. The power dissipation is only $\leq2\times10^{-16}$ Watts, ruling out the possibility of Joule-heating. Within a current range of $\pm 2.5 p$A, the $V-I$ is linear and 
corresponds to a resistance of $74$ M$\Omega$ which is 200 times more than the EHS case. The threshold is reached at  $I_{drive}\sim2.5p$A right at the end of the linear range beyond which a fifteen-fold decrease in the differential resistance $dV/dI$ appears [Fig.~\ref{fig:29mK}(b)]. This rising pattern of resistivity ($\rho$) exhibits a striking difference than the EHS case~\cite{WC_helium} in which $\rho$ increases nonlinearly first before reaching the threshold. Yet, it is not entirely surprising considering a different disorder pinning mechanism for ($I\leq I_c$) than the EHS case where pinning is via ripplon drag~\cite{ripplon}.

The effect of disorder, especially due to long-ranged charge impurities, probably has the most profound impact to the dynamical pinning of a WC. Distortions and even local phase transitions, as simulated for a classical WC~\cite{nori}, could occur with increasing current drive and gives rise to a complicated transport. However, the absence of hysteresis above and below $I_c$ as shown in Fig.\ref{fig:29mK}(a) provides a clue for an intact WC. Another clue is found through the ``wiggles" around the threshold which are better seen as slight oscillations in the inset zoom-in view. This is impossible from a highly defected WC which, under bias, form a network of winding channels~\cite{nori}. Instead, such oscillations are consistent with the presence of discrete edge channels.


The 2D charge distribution is uniform everywhere except at sample edges where $p$ must diminish. This density gradient at the edge leads to the formation of anisotropic discrete edge microchannels/filaments~\cite{Bajaj} that dynamically decouple with the WC and vary with the driving and pinning forces. As $I_{drive}$ increases, as also seen for EHS~\cite{WC_helium}, the carriers at the edge melt even when $I<I_c$ and the number of the edge filaments increases. Voltage oscillations result from the switching on of new filaments. Further evidence will be seen later with the T-dependence results. However, the transport through the edge channels is not sufficient to explain the linear resistance below $I_C$ because the oscillations are too small compared to the overall resistance.

Considering the highly suppressed disorder level, the $74$ M$\Omega$ linear resistance indicates an enormous pinning for which one might also anticipate a large depinning threshold. However, the remarkably small $I_C\sim 2.5 p$A is only $\sim1/10^3$ of that found in EHS and $1/10^6$ of that for the CDW cases. A more rigorous approach is needed because the threshold in not only determined by disorder, but also the depinning mechanism of a quantum WC, even with possible quantum mechanical mechanisms~\cite{Bardeen} instead of just the classical sliding of charges. For example, the absence of hysteresis is consistent with a quantum effect rather than a classical one~\cite{CDWreview}.

Nevertheless, a comparison is made to an available theory for a classical WC. The switching region of the $V-I$ is compared to a power law $I-I_c=\alpha(V-V_c)^{\xi}$~\cite{nori}. The free parameter fitting, shown as the solid line in Fig.~\ref{fig:29mK}, yields $V_c\sim 167\mu$V, $I_c\sim2.35p$A, and $\alpha\sim1.7\times10^{-4}$. The exponent $\xi$ is $\simeq1.85$, similar to previous findings.\\


\begin{figure}
\vspace{-0pt}
\includegraphics[totalheight=5.4in,trim=0.35in 0.10in 0.20in 0in]{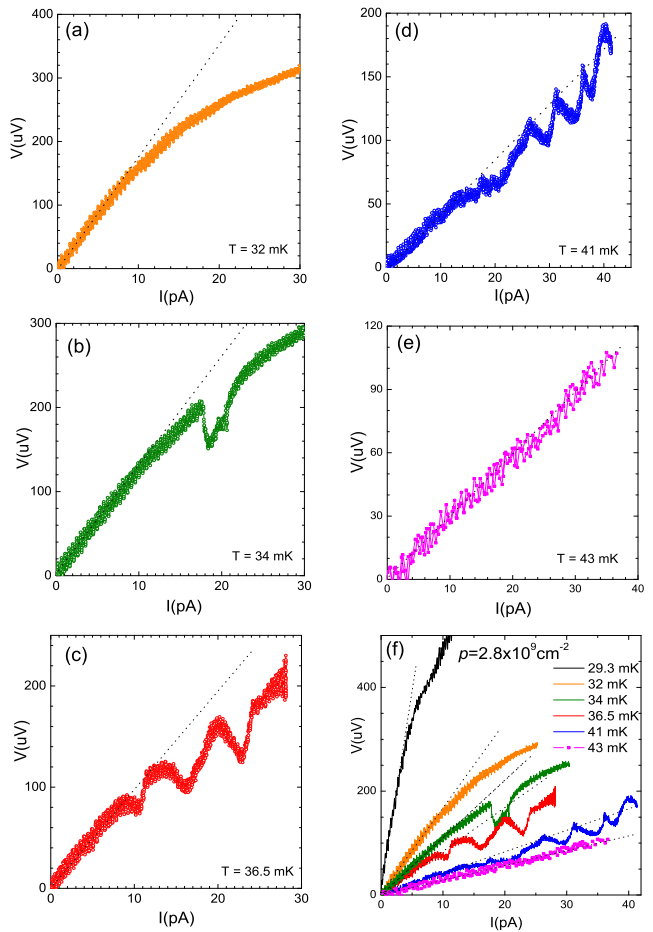}
\vspace{-0pt}
\caption{\label{fig:dc} (Color online) The dc $I-V$ results for various temperatures (a) 32 mK, (b) 34 mK, (c) 36.5 mK, (d) 41 mK, (e) 43 mK.  All $V-I$s (including Fig.~\ref{fig:29mK}) are compared in (f).}
\vspace{-0pt}
\end{figure}

Below we present the T-dependence of the dc VI characteristics for which little has been reported on this important subject. Fixing at the same charge density, we increase T to several fixed values at which the same dc $V-I$ measurement is repeated. First, the T-dependence of the differential resistance $dV/dI$ derived from each of the linear portions of the ($V-I$)s shows a nonactivated transport as shown in a log-log plot in figure~\ref{fig:29mK}(c). This is consistent with the ac results~\cite{jian-image} and other HIGFET measurements~\cite{jian-1} which confirms that the system is not an Anderson insulator~\cite{jian-image,jian-1}. This is an important point since an Anderson insulator can also produces nonlinear $V-I$. The voltage oscillation is also an evidence against the presence of an Anderson insulator because the overall measured signal must yield an averaged effect which smooths out individual voltage jumps from random single particle hoppings.

The $V-I$ results are shown in figure~\ref{fig:dc}. The 32 mK result shown in (a) is similar to the 29.3 mK result with the same slight oscillation around the threshold. The linear window is slightly larger, yielding an increased threshold to $\sim4.6 p$A even though the bending itself is somewhat weaker. The corresponding resistivity $\rho$ for the linear portion is 22 M$\Omega$, approximately 1/4 of that for $T$=29.3 mK for 3mK temperature rise. Upon reaching 34 mK, the same nonlinear behavior is observed with the threshold moved to $\sim6.3 p$A. However, the "wiggle" is replaced with a significant dip around $I=18 p$A, with a substantial negative differential resistance (NDR). The measured voltage $V$ drops 55 $\mu$V which is 25\% of the whole magnitude over a current increase of $0.8 p$A, before it is recovered at $I\simeq24 p$A. Upon reaching $T=36.5 m$K, as shown in (c), an oscillatory $V-I$ is clearly developed just above the threshold, with the dips/minimums almost evenly spaced at 10.3, 16.2, and 22.7 $p$A. NDR appears three times between 9 and 23 $p$A. The amplitudes of the oscillations are approximately 25\%. The threshold $I_c$ is $\sim8.3 p$A. The nonlinear bending appears even weaker. As T is further increased to $T=41 m$K [shown in (d)], the $V-I$ oscillations are seen for higher currents yet with slightly reduced amplitudes around 25 $\mu$V. The nonlinear threshold feature is further weakened. Upon reaching 43 mK [(e)], the oscillations diminish and the $V-I$ already becomes approximately linear in the same current range. All $V-I$ curves are plotted in (f) for comparison.

\begin{figure}
\vspace{-0pt}
\includegraphics[totalheight=2.1in,trim=0.2in 0.10in 0.20in 0in]{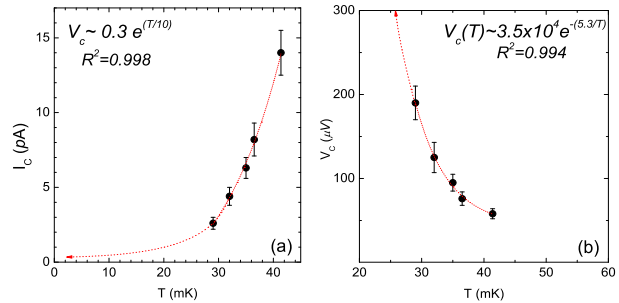}
\vspace{-0pt}
\caption{\label{fig:R} T-dependence of (a) the threshold current $I_c$ and (b) the threshold voltage. (dotted lines are exponential fits).}
\vspace{-0pt}
\end{figure}

The increase of T weakens the WC by raising the energy closer to the melting point $T_m$. Consequently, disorder pinning is effectively reduced as the rigidity of the WC becomes less. This is consistent with the decrease of the resistivity$\rho$ (in the linear region) with increasing T, as well as the increase in $I_c(T)$. As plotted in figure~\ref{fig:R}(a), the growth of $I_c(T)$ can be fitted to $\sim 0.3e^{T/10}$ (with $R^2=0.998$) shown as a dotted line. The corresponding voltage is shown in panel (b) which is also fitted to an exponential function. $I_c$ becomes hard to resolve as T goes above 36 mK and disappears when T$>41$ mK. One possibility is that the WC is melted which is well supported by the T-dependence of the voltage oscillations. The amplitudes of the voltage oscillations first increase due to the widening of the edge channels and the rapid increase of the number of filaments with increasing $I_{drive}$. Then, as the WC starts to melt, the amplitudes of the oscillations decrease and eventually disappear at T$>41$ mK when it becomes a Wigner liquid. This also explains why the results from Ref.~\cite{Yoon} for a GaAs system at T above 50 mK did not exhibit any voltage oscillations. This lower $T_m$ for this quantum WC case, which is approximately one-third the classical estimate of $\sim E_C/137\sim 120 m$ K, can be explained by combining the effects of quantum fluctuations~\cite{Glazman}, residual disorder~\cite{nori}, and the current excitation used for the measurement.

The modeling for the depinning of a WC against Coulomb disorder, especially with possible quantum mechanical processes, is beyond the scope of this discussion and more theoretical work is needed. Nevertheless, there are clearly striking features that are inconsistent with a classical sliding charge picture. First, $I_c(T)$ is weakly T-dependent at low-T and extrapolates to a finite value in the limit of $T\rightarrow 0$ when the role of quantum effects are important, suggesting a quantum nature of $I_c$. This is consistent with the overwhelmingly smaller $I_c$ than all classical cases. Second, since the classical pinning is a nonliner effect, $\rho$, as for the EHS~\cite{WC_helium} and the Si-MOSFET~\cite{Kravchenko} cases, should rise nonlinearly due to the rising pinning force before reaching the threshold. This characteristic nonlinear rise is, however, absent in our case as shown in both Fig.~\ref{fig:29mK} and Fig.~\ref{fig:dc}. Quantum processes, such as tunneling, have a better chance of explaining it. Third, hysteretic behaviors, as already shown for the classical scenario~\cite{CDWreview}, is a slow process that is incompatible with quantum processes. In our case, no sign of hysteresis is found.

The quantum depinning was first considered theoretically~\cite{Bardeen} for CDW studies. There, a threshold is set by Coulomb blockade energy associated with an internal electric field generated through tunneling events~\cite{Maki,Rozhavsky,Miller}. This notion can be better tested in a quantum WC case in which tunneling through a narrow bandgap is primarily influenced by the energy of the carriers. Therefore, increasing the driving field continuously improves the tunneling probability and there is no need for the increase of resistivity all the way up to the (quantum) depinning threshold when the manybody carriers have sufficient energy to tunnel at large scales. This idea motivates a dc+ac measurement close to the threshold.


To summarize, through measuring an ultra-dilute 2D hole system in GaAs HIGFET at extremely small excitations in zero magnetic field, we have observed a threshold transport behavior along with resistance oscillations which are the characteristics for the pinning and depinning of a Wigner crystal. Raising the driving current leads to a nonequilibrium situation in which the current, due to both the moving crystal and the ordered edge filaments, coexist. Moreover, multiple facts, such as the remarkably small $I_c$, the melting of the WC well below the classical point, the absence of hysteresis and nonlinear increase of resistivity, etc are not compatible with the classical depinning picture and maybe indications for a quantum depinning mechanism. Many questions have been left unanswered at this juncture. Especially, the question on how to establish a distinction between the classical and the quantum pinning/depinning mechanisms awaits further exploration.

We acknowledge the support of this work from NSF under DMR-1105183. The work at Princeton was partially funded by the Gordon and Betty Moore Foundation through Grant GBMF2719, and by the National Science Foundation MRSEC-DMR-0819860 at the Princeton Center for Complex Materials.



\end{document}